\documentclass{jpsj2}
\title{Location of the Multicritical Point for the Ising Spin Glasses
on the Triangular and Hexagonal Lattices}

\author{Hidetoshi Nishimori and Masayuki Ohzeki}

\inst{Department of Physics, Tokyo Institute of Technology,
Oh-okayama, Meguro-ku, Tokyo 152-8551}

\abst{A conjecture is given for the exact location of the multicritical
point in the phase diagram of the $\pm J$ Ising model on the triangular
lattice.
The result $p_{\rm c}=0.8358058$ agrees well with a recent numerical estimate.
From this value, it is possible to derive a comparable conjecture for
the exact location of the multicritical point for the
hexagonal lattice, $p_{\rm c}=0.9327041$, again in excellent agreement
with a numerical study.
The method is a variant of duality transformation to relate the
triangular lattice
directly with its dual triangular lattice without recourse to the
hexagonal lattice, in conjunction with the replica method.
}

\kword{Spin glass, $\pm J$ Ising model, multicritical point, exact solution,
triangular lattice, hexagonal lattice, duality transformation}
\begin{document}
\maketitle
\section{Introduction}
Properties of finite-dimensional spin glasses are still under active current
investigations after thirty years since the proposal of basic models
\cite{EA,SK}.
Among issues are the existence and critical properties of spin glass
transition, low-temperature slow dynamics, and competition between the
spin glass and conventional phases \cite{Young}.

Determination of the structure of the phase diagram belongs to this last 
class of problems.
Recent developments of analytical theory for this purpose
\cite{NN,MNN,TN,TSN}, namely a
conjecture on the exact location of the multicritical point, have opened
a new perspective.
The exact value of the multicritical point, in addition to its intrinsic interest
as one of the rare exact results for finite-dimensional spin glasses, greatly
facilitates precise determination of critical exponents around the multicritical
point in numerical studies.

The theory to derive the conjectured exact location of the multicritical point
has made use of the replica method in conjunction with duality transformation.
The latter aspect restricts the direct application of the method to self-dual lattices.
It has not been possible to predict the location of the multicritical point
for systems on the triangular and hexagonal lattices although a relation
between these mutually-dual cases has been given \cite{TSN}.

In the present paper we use a variant of duality transformation to derive
a conjecture for the $\pm J$ Ising model on the triangular lattice.
The present type of duality allows us to directly map the triangular lattice
to another triangular lattice without recourse to the hexagonal lattice
or the star-triangle transformation adopted in the conventional approach.
This aspect is particularly useful to treat the present disordered system under
the replica formalism as will be shown below.
The result agrees impressively with a recent numerical estimate of high precision.
This lends additional support to the reliability of our theoretical
framework \cite{NN,MNN,TN,TSN} to derive a conjecture on the exact location of the
multicritical point for models of finite-dimensional spin glasses.

\section{Ferromagnetic system on the triangular lattice}

It will be useful to first review the duality transformation for
the non-random $Z_q$ model on the triangular lattice formulated
without explicit recourse to the hexagonal lattice or the star-triangle
transformation \cite{Wu}.
Let us consider the $Z_q$ model with an edge Boltzmann factor
$x[\phi_i-\phi_j]$ for neighbouring sites $i$ and $j$.
The spin variables $\phi_i$ and $\phi_j$ take values
from 0 to $q-1$ (mod $q$). The function $x[\cdot]$ itself is
also defined with mod $q$.
An example is the clock model with coupling $K$,
\begin{equation}
  x[\phi_i-\phi_j]=\exp \left\{ K\cos \left(\frac{2\pi }{q}
  (\phi_i-\phi_j) \right)\right\}.
\end{equation}
The Ising model corresponds to the case $q=2$.

\begin{figure}[tb]
\begin{center}
\includegraphics[width=60mm]{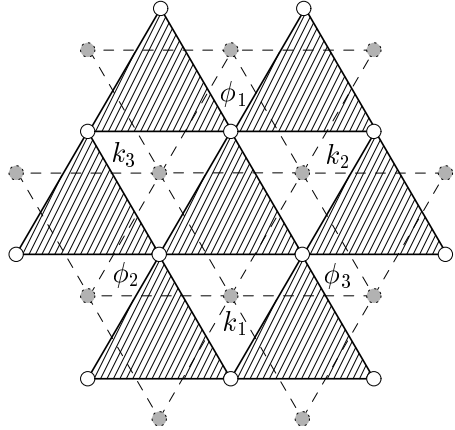}
\end{center}
\caption{Triangular lattice and its dual (shown dashed).
The up-pointing triangle surrounded by the variables $\phi_1, \phi_2, \phi_3$
is transformed into the down-pointing triangle surrounded by $k_1, k_2, k_3$.}
\label{fig1}
\end{figure}

The partition function may be written as
\begin{equation}
 Z=q \sum_{\{ \phi_{ij}\}}
 \prod_{\bigtriangleup}x[\phi_{12}]x[\phi_{23}]x[\phi_{31}]
 \delta (\phi_{12}+\phi_{23}+\phi_{31})
 \prod_{\bigtriangledown}\delta \left(\sum \phi_{ij}\right).
\label{Zdef}
\end{equation}
Here the product over $\bigtriangleup$ runs over up-pointing triangles
shown shaded in Fig. \ref{fig1} and that for $\bigtriangledown$
is over unshaded down-pointing triangles.
The variable of summation is not written as the original $\phi_i$
but in terms of the directed difference $\phi_{ij}=\phi_i-\phi_j$
defined on each bond.
This is possible if we introduce restrictions represented by
the Kronecker deltas (which are defined with mod $q$) as in eq. (\ref{Zdef})
allocated to all up-pointing and down-pointing triangles.
For instance, $\phi_{12}(=\phi_1-\phi_2)$, $\phi_{23}(=\phi_2-\phi_3)$,
and $\phi_{31}(=\phi_3-\phi_1)$ are not independent but satisfy
$\phi_{12}+\phi_{23}+\phi_{31}=0$ (mod $q$), where 1, 2, and 3 are sites around
the unit triangle as indicated in Fig. \ref{fig1}.
The overall factor $q$ on the right hand side of eq. (\ref{Zdef}) reflects
the invariance of the system under the uniform change
$\phi_i\to\phi_i+l~(\forall i, 0\le l \le q-1)$.

It is convenient to Fourier-transform the Kronecker deltas
for down-pointing triangles and allocate the resulting exponential
factors to the edges of three neighbouring up-pointing triangles.
Then the partition function can be written only in terms of a product
over up-pointing triangles:
\begin{equation}
 Z=q \sum_{\{k_i\}}\sum_{\{ \phi_{ij}\}}
 \prod_{\bigtriangleup}\left[\frac{1}{q}A[\phi_{12}, \phi_{23}, \phi_{31}]
 \delta (\phi_{12}+\phi_{23}+\phi_{31})
 \exp\left\{ \frac{2\pi i}{q}(k_1\phi_{23}+k_2\phi_{31}+k_3\phi_{12})\right\}\right],
\label{Z2}
\end{equation}
where $A[\phi_{12}, \phi_{23}, \phi_{31}]=x[\phi_{12}]x[\phi_{23}]x[\phi_{31}]$.

Now let us regard
the product over up-pointing triangles in eq. (\ref{Z2})
as a product over down-pointing triangles overlaying the original up-pointing
triangles as shown dashed in Fig. \ref{fig1}.
This viewpoint allows us to regard the quantity in the square brackets of
eq. (\ref{Z2}) as the Boltzmann factor for the unit triangle (to be called the
face Boltzmann factor hereafter)
of the dual triangular lattice composed of overlaying down-pointing triangles:
\begin{equation}
 Z=q \sum_{\{k_i\}}
 \prod_{\bigtriangledown^*}A^*[k_{12}, k_{23}, k_{31}],
\end{equation}
where
\begin{eqnarray}
   A^*[k_{12}, k_{23}, k_{31}]&=&\frac{1}{q}\sum_{\phi_{12},\phi_{23},\phi_{31}=0}^{q-1}
   A[\phi_{12},\phi_{23},\phi_{31}]\delta (\phi_{12}+\phi_{23}+\phi_{31})
  \nonumber\\
 &&\times\exp\left\{ \frac{2\pi i}{q}
(k_1\phi_{23}+k_2\phi_{31}+k_3\phi_{12})\right\}.
\label{A_dual}
\end{eqnarray}
Here we have used the fact that the right hand side is a function of the
differences $k_i -k_j\equiv k_{ij}~((ij)=(12), (23), (31))$ due to the constraint
$\phi_{12}+\phi_{23}+\phi_{31}=0$.

This is a duality relation which exchanges the original model on the triangular
lattice with a dual system on the dual triangular lattice.
$A^*[k_{12},k_{23},k_{31}]$ represents the face Boltzmann factor of the dual system, which is the function of the differences between the nearest neighbor sites on the unit triangles similarly to face Boltzmann factor $A[\phi_{12},\phi_{23},\phi_{31}]$ of the original system. 

It is easy to verify that the usual duality relation for the triangular lattice
emerges from the present formulation.  As an example, the ferromagnetic
Ising model on the triangular lattice has the following face Boltzmann factors:
\begin{equation}
 A[0, 0, 0]={\rm e}^{3K},~~A[1, 1, 0]=A[1,0,1]=A[0,1,1]={\rm e}^{-K},\cdots ,
\end{equation}
where $A[0, 0, 0]$ is for the all-parallel neighbouring spin configuration
for three edges of a unit triangle,
and $A[1, 1, 0]$ is for
two antiparallel pairs and a single parallel pair around a unit triangle.
The dual are, according to eq. (\ref{A_dual}),
\begin{equation}
A^*[0,0,0]=\frac{1}{2}\left\{A[0,0,0]+3A[1,1,0]\right\},
~A^*[1,1,0]=\frac{1}{2}\left\{A[0,0,0]-A[1,1,0]\right\}.
\end{equation}
It then follows that
\begin{equation}
  {\rm e}^{-4K^*}\equiv \frac{A^*[1,1,0]}{A^*[0,0,0]}
 =\frac{1-{\rm e}^{-4K}}{1+3{\rm e}^{-4K}}.
\end{equation}
This formula is equivalent to the expression obtained by the ordinary duality,
which relates the triangular lattice to the hexagonal lattice,
followed by the star-triangle transformation:
\begin{equation}
  (1+3{\rm e}^{-4K})(1+3{\rm e}^{-4K^*})=4.
\end{equation}
%
\section{Replicated system}
It is straightforward to generalize the formulation of the previous
section to the spin glass model using the replica method.
The duality relation for the face Boltzmann factor of the replicated system is
\begin{eqnarray}
  A^*[\{k_{12}^{\alpha}\},\{k_{23}^{\alpha}\},\{k_{31}^{\alpha}\}]
 &=&\frac{1}{q^n}
 \sum_{\{\phi\}} \left[ \prod_{\alpha=1}^n
 \delta (\phi_a^{\alpha}+\phi_b^{\alpha}+\phi_c^{\alpha})\right]
  A\left[\{\phi_a^{\alpha}\},\{\phi_b^{\alpha}\},\{\phi_c^{\alpha}\}\right]
   \nonumber\\
 &&\times\exp \left\{\frac{2\pi i}{q}
 \sum_{\alpha=1}^n (k_1^{\alpha}\phi_b^{\alpha}+k_2^{\alpha}\phi_c^{\alpha}
  +k_3^{\alpha}\phi_a^{\alpha}) \right\},
  \label{Astar}
\end{eqnarray}
where $\alpha$ is the replica index running from 1 to $n$,
$\{k_{ij}^{\alpha}\}$ denotes the set $\{k_{ij}^{1}, k_{ij}^2, \cdots, k_{ij}^n\}$,
and similarly for $\{\phi_a^{\alpha}\}$ etc.
The variables $\phi_a^{\alpha}, \phi_b^{\alpha}, \phi_c^{\alpha}$ correspond
to $\phi_{12},\phi_{23},\phi_{31}$ in eq. (\ref{Z2}).
The original face Boltzmann factor is the product of three edge Boltzmann
factors
\begin{equation}
  A\left[\{\phi_a^{\alpha}\},\{\phi_b^{\alpha}\},\{\phi_c^{\alpha}\}\right]
 =\chi_{\phi_a^1\cdots \phi_a^n}\cdot\chi_{\phi_b^1\cdots \phi_b^n}\cdot
 \chi_{\phi_c^1\cdots \phi_c^n},
\end{equation}
where $\chi_{\phi^1\cdots \phi^n}$ is the averaged edge Boltzmann factor
\begin{equation}
 \chi_{\phi^1\cdots \phi^n}=\sum_{l=0}^{q-1}p_l \, x[\phi^1+l]x[\phi^2+l]\cdots
   x[\phi^n+l].
\end{equation}
Here $p_l$ is the probability that the relative value of neighbouring spin
variables is shifted by $l$.  A simple example is the $\pm J$ Ising model
($q=2$), in which $p_0=p$ (ferromagnetic interaction) and $p_1=1-p$
(antiferromagnetic interaction).

The average of the replicated partition function $Z_n$ is a function of
face Boltzmann factors for various values of $\phi$'s.
The triangular-triangular duality relation is then written as\cite{MNN,TN,TSN}
\begin{equation}
 Z_n(A[\{0\},\{0\},\{0\}],\cdots )
 =cZ_n(A^*[\{0\},\{0\},\{0\}],\cdots ),
 \label{duality_Zn}
\end{equation}
where $c$ is a trivial constant and $\{0\}$ denotes the set of $n$ 0's.
Since eq. (\ref{duality_Zn}) is a duality relation for a multivariable function,
it is in general impossible to identify the singularity of the system with
the fixed point of the duality transformation.
Nevertheless, it has been firmly established in simpler cases (such as the square lattice)
that the location of the multicritical point in the phase diagram of spin glasses
can be predicted very accurately, possibly exactly, by using the fixed-point
condition of the principal Boltzmann factor for all-parallel configuration $\{0\}$.
\cite{NN,MNN,TN,TSN}
We therefore try the ansatz also for the triangular lattice that the exact
location of the multicritical
point of the replicated system is given by the fixed-point condition
of the principal face Boltzmann factor:
\begin{equation}
 A\left[\{0\},\{0\},\{0\}\right]=A^*\left[\{0\},\{0\},\{0\}\right],
 \label{AAstar}
\end{equation}
combined with the Nishimori line (NL) condition, on which the multicritical
point is expected to lie \cite{HN81,HNbook}.

For simplicity, we restrict ourselves to the $\pm J$ Ising model hereafter.
Then the NL condition is ${\rm e}^{-2K}=(1-p)/p$, where
$p$ is the probability of ferromagnetic interaction.
The original face Boltzmann factor is a simple product of three edge Boltzmann
factors,
\begin{equation}
  A\left[\{0\},\{0\},\{0\}\right]=\chi_0^3
\end{equation}
with
\begin{equation}
 \chi_0=p{\rm e}^{nK}+(1-p){\rm e}^{-nK}=\frac{{\rm e}^{(n+1)K}
+{\rm e}^{-(n+1)K}}{{\rm e}^K+{\rm e}^{-K}}.
\end{equation}

The dual Boltzmann factor $A^*[\{0\},\{0\},\{0\}]$ needs a more
elaborate treatment.
The constraint of Kronecker delta in eq. (\ref{Astar}) may be expressed as
\begin{equation}
  \prod_{\alpha=1}^n
 \delta (\phi_a^{\alpha}+\phi_b^{\alpha}+\phi_c^{\alpha})
 =2^{-n}\sum_{\{\tau_{\alpha}=0,1\}}\exp\left[ \pi i\sum_{\alpha =1}^n \tau_{\alpha}
  (\phi_a^{\alpha}+\phi_b^{\alpha}+\phi_c^{\alpha})\right].
 \label{product_delta}
\end{equation}
The face Boltzmann factor
$A\left[\{\phi_a^{\alpha}\},\{\phi_b^{\alpha}\},\{\phi_c^{\alpha}\}\right]$
is the product of three edge Boltzmann factors, each of which may be written
as, on the NL, \cite{TSN}
\begin{eqnarray}
  \chi_{\phi_a^1\cdots \phi_a^n}
  &=&p \exp\left[ \sum_{\alpha=1}^n (1-2\phi_a^{\alpha})K\right]+
  (1-p)\exp\left[-\sum_{\alpha=1}^n (1-2\phi_a^{\alpha})K\right]
 \nonumber\\
 &=&\frac{1}{2\cosh K}\sum_{\eta_a=\pm 1}\exp\left[\eta_a K+\eta_a K
  \sum_{\alpha=1}^n (1-2\phi_a^{\alpha}K)\right].
  \label{chi_a}
\end{eqnarray}
Using eqs. (\ref{product_delta}) and (\ref{chi_a}), eq. (\ref{Astar}) can be
rewritten as, for the principal Boltzmann factor with all $k_i^{\alpha}=0$,
\begin{eqnarray}
 && A^*\left[\{0\},\{0\},\{0\}\right]
  =\frac{1}{4^n (2\cosh K)^3} \sum_{\eta}\sum_{\tau}\sum_{\phi}
    \exp\left[ \pi i \sum_{\alpha}\tau_{\alpha}(\phi_a^{\alpha}
      +\phi_b^{\alpha}+\phi_c^{\alpha})\right.
      \nonumber\\
    &&+\left. K(\eta_a+\eta_b+\eta_c)+K\eta_a\sum_{\alpha}(1-2\phi_a^{\alpha})
      +K\eta_b\sum_{\alpha}(1-2\phi_b^{\alpha})
      +K\eta_c\sum_{\alpha}(1-2\phi_c^{\alpha})\right].
      \label{AAexp}
 \end{eqnarray}
The right hand side of this equation can be evaluated explicitly as shown in the Appendix.
The result is
 \begin{eqnarray}
   &&A^*\left[\{0\},\{0\},\{0\}\right]=4^{-n}(2\cosh K)^{3n-3}
   \nonumber\\
  && \times \left[ ({\rm e}^{3K}+3{\rm e}^{-K})(1+\tanh^3 K)^n
     +(3{\rm e}^{K}+{\rm e}^{-3K})(1-\tanh^3 K)^n \right].
     \label{AAresult}
 \end{eqnarray}
The prescription (\ref{AAstar}) for the multicritical point is therefore
\begin{eqnarray}
  &&\frac{\cosh^3 (n+1)K}{\cosh^3 K}=
  \frac{(2\cosh K)^{3n}}{4^n (2\cosh K)^3} 
      \nonumber\\
    &&\times \left[ ({\rm e}^{3K}+3{\rm e}^{-K})(1+\tanh^3 K)^n
     +(3{\rm e}^{K}+{\rm e}^{-3K})(1-\tanh^3 K)^n \right].
      \label{MCP}
 \end{eqnarray}
%

\section{Multicritical point}
The conjecture (\ref{MCP}) for the exact location of the multicritical point can
be verified for $n=1, 2$ and $\infty$ since these cases can be treated
directly without using the above formulation.

The case $n=1$ is an annealed system and the problem can be solved explicitly.
It is easy to show that the annealed $\pm J$ Ising model is equivalent to
the ferromagnetic Ising model with effective coupling $\tilde{K}$ satisfying
\begin{equation}
  \tanh \tilde{K}=(2p-1)\tanh K.
  \label{annealed1}
\end{equation}
If we insert the transition point of the ferromagnetic Ising model
on the triangular lattice ${\rm e}^{4\tilde{K}}=3$, eq. (\ref{annealed1})
reads
\begin{equation}
 (2p-1)\tanh K=2-\sqrt{3}.
 \label{annealed2}
\end{equation}
This formula represents the exact phase boundary for the annealed system.
Under the NL condition, it is straightforward to verify that this expression
agrees with the conjectured multicritical point of eq. (\ref{MCP}).
It is indeed possible to show further that the whole phase boundary
of eq. (\ref{annealed2}) can be derived by evaluating $A^{*}[0,0,0]$
directly for $n=1$ for arbitrary $p$ and $K$,
giving $2A^{*}[0,0,0]=\chi_0^3+3\chi_1^2 \chi_0$ with
$\chi_0=p{\rm e}^K+(1-p){\rm e}^{-K}, \chi_1=p{\rm e}^{-K}+(1-p){\rm e}^{K}$,
and using the condition $A^{*}[0,0,0]=A[0,0,0]$.

When $n=2$, a direct evaluation of the edge Boltzmann factor reveals that
the system on the NL is a four-state Potts model with effective
coupling $\tilde{K}$ satisfying \cite{TSN,MNN}
\begin{equation}
  {\rm e}^{\tilde{K}}={\rm e}^{2K}-1+{\rm e}^{-2K}.
\end{equation}
Since the transition point of the non-random four-state Potts model on the triangular
lattice is given by ${\rm e}^{\tilde{K}}=2$ \cite{Wu}, the (multi)critical
point
of the $n=2$ system is specified by the relation
\begin{equation}
  {\rm e}^{2K}-1+{\rm e}^{-2K}=2.
\end{equation}
Equation (\ref{MCP}) with $n=2$ also gives this same expression, which
confirms validity of our conjecture in the present case as well.

The limit $n\to\infty$ can be analyzed as follows.\cite{MNN}
The average of the replicated partition function
\begin{equation}
 [Z^n]_{\rm av}=[{\rm e}^{-n\beta F}]_{\rm av},
\end{equation}
where $[\cdots ]_{\rm av}$ denotes the configurational average,
is dominated in the limit $n\to\infty$ by contributions from bond
configurations with the smallest value of the free energy $F$.
It is expected that the bond configurations without frustration
(i.e., ferromagnetic Ising model and its gauge equivalents) have the
smallest free energy, and therefore we may reasonably expect that the $n\to\infty$
systems is described by the non-random model.
Thus the critical point is given by ${\rm e}^{4K}=3$.
It is straightforward to check that eq. (\ref{MCP}) reduces
to the same equation in the limit $n\to\infty$.

These analyses give us a good motivation to apply eq. (\ref{MCP})
to the quenched limit $n\to 0$.
Expanding eq. (\ref{MCP}) around $n=0$, we find,
from the coefficients of terms linear in $n$,
\begin{eqnarray}
 3K\tanh K&=&3 \log (2\cosh K)-\log 4
  \nonumber\\
  &&+\frac{{\rm e}^{3K}+3{\rm e}^{-K}}{(2\cosh K)^3}\log (1+\tanh^3 K)
  +\frac{3{\rm e}^{K}+{\rm e}^{-3K}}{(2\cosh K)^3}\log (1-\tanh^3 K),
\end{eqnarray}
or, in terms of $p$, using ${\rm e}^{-2K}=(1-p)/p$,
\begin{eqnarray}
  &&2p^2(3-2p)\log p+2(1-p)^2(1+2p)\log (1-p)+\log 2
   \nonumber\\
   && =p(4p^2-6p+3)\log (4p^2-6p+3)
     +(1-p)(4p^2-2p+1)\log (4p^2-2p+1).
     \label{pc}
\end{eqnarray}
Equation (\ref{pc}) is our conjecture for the exact location of the
multicritical point $p_{\rm c}$
of the $\pm J$ Ising model on the triangular lattice.
This gives $p_{\rm c}=0.8358058$, which agrees well with a recent
high-precision numerical estimate, 0.8355(5) \cite{Queiroz}.

If we further use the conjecture \cite{TSN} $H(p_{\rm c})+H(p_{\rm c}')=1$,
where $H(p)$ is the binary entropy $-p\log_2 p-(1-p)\log_2 (1-p)$, to relate this
$p_{\rm c}$ with that for the hexagonal lattice
$p_{\rm c}'$, we find $p_{\rm c}'=0.9327041$.
Again, the numerical result 0.9325(5) \cite{Queiroz} is very close 
to this conclusion.

\section{Conclusion}

To summarize, we have formulated the duality transformation of the replicated
random system on the triangular lattice, which brings the triangular lattice
to a dual triangular lattice without recourse to the hexagonal lattice.
The result was used to predict the exact location of the multicritical point
of the $\pm J$ Ising model on the triangular lattice.
Correctness of our theory has been confirmed in directly solvable cases
of $n=1, 2$ and $\infty$.
Application to the quenched limit $n\to 0$ yielded a value in impressive
agreement with a numerical estimate.

The status of our result for the quenched system, eq. (\ref{pc}), is
a conjecture for the exact solution.
It is difficult at present to prove this formula rigorously.
This is the same situation as in cases for other lattices and models
\cite{NN,MNN,TN,TSN}.
We nevertheless expect that such a proof should be eventually possible
since a single unified theoretical framework
always gives results in excellent agreement with independent numerical
estimations for a wide range of systems.
Further efforts toward a formal proof are required.
\section*{Acknowledgement}
This work was supported by the Grant-in-Aid for Scientific Research on
Priority Area ``Statistical-Mechanical Approach to Probabilistic Information
Processing" by the MEXT.

\section*{Appendix}
In this Appendix we evaluate
eq. (\ref{AAexp}) to give eq. (\ref{AAresult}).
Let us denote $4^n (2\cosh K)^3A^*[\{0\},\{0\},\{0\}]$ as $\tilde{A}$.
The sums over $\alpha$ in the exponent of eq. (\ref{AAexp}) can be expressed
as the product over $\alpha$:
\begin{eqnarray}
  \tilde{A}&=&\sum_{\eta}{\rm e}^{K(n+1)(\eta_a+\eta_b+\eta_c)}
   \nonumber\\
  &\times& \prod_{\alpha=1}^n \left[
     \sum_{\tau_{\alpha}=0}^{1}\left(
     \sum_{\phi_{a}^{\alpha}=0}^1
     {\rm e}^{\pi i\tau_{\alpha}\phi_a^{\alpha}-2K\eta_a \phi_a^{\alpha}}
    \sum_{\phi_{b}^{\alpha}=0}^1
     {\rm e}^{\pi i\tau_{\alpha}\phi_b^{\alpha}-2K\eta_b \phi_b^{\alpha}}
     \sum_{\phi_{c}^{\alpha}=0}^1
     {\rm e}^{\pi i\tau_{\alpha}\phi_c^{\alpha}-2K\eta_c \phi_c^{\alpha}}\right)
    \right].
\end{eqnarray}
By performing the sums over $\phi$ and $\tau$ for each replica, we find
\begin{eqnarray}
 \tilde{A}&=&\sum_{\eta_a,\eta_b,\eta_c=\pm 1}
   {\rm e}^{K(n+1)(\eta_a+\eta_b+\eta_c)}
 \prod_{\alpha=1}^n \left[
(1+{\rm e}^{-2K\eta_a})(1+{\rm e}^{-2K\eta_b})(1+{\rm e}^{-2K\eta_c})\right.
\nonumber\\
&& \left.
+(1-{\rm e}^{-2K\eta_a})(1-{\rm e}^{-2K\eta_b})(1-{\rm e}^{-2K\eta_c})\right].
\end{eqnarray}
It is straightforward to write down the eight terms appearing in the
above sum over $\eta_a, \eta_b, \eta_c$ to yield
\begin{eqnarray}
  \tilde{A}&=&{\rm e}^{3K(n+1)}\left[
 (1+{\rm e}^{-2K})^3+(1-{\rm e}^{-2K})^3\right]^n
  \nonumber\\
&+&3{\rm e}^{K(n+1)}\left[(1+{\rm e}^{-2K})^2 
(1+{\rm e}^{2K})+(1-{\rm e}^{-2K})^2(1-{\rm e}^{2K})\right]^n
 \nonumber\\
&+&3{\rm e}^{-K(n+1)}\left[(1+{\rm e}^{-2K})
(1+{\rm e}^{2K})^2+(1-{\rm e}^{-2K})(1-{\rm e}^{2K})^2\right]^n
 \nonumber\\
&+&{\rm e}^{-3K(n+1)}\left[(1+{\rm e}^{2K})^3+(1-{\rm e}^{2K})^3\right]^n,
\end{eqnarray}
which is further simplified into
\begin{eqnarray}
  \tilde{A}&=&{\rm e}^{3K}\left[(2\cosh K)^3+(2\sinh K)^3\right]^n
\nonumber\\
 &+&3{\rm e}^{K}\left[(2\cosh K)^3-(2\sinh K)^3\right]^n
\nonumber\\
 &+&3{\rm e}^{-K}\left[(2\cosh K)^3+(2\sinh K)^3\right]^n
 \nonumber\\
 &+&{\rm e}^{-3K}\left[(2\cosh K)^3-(2\sinh K)^3\right]^n
\nonumber\\
  &=& (2\cosh K)^{3n}
 \nonumber\\
 &&\times\left[({\rm e}^{3K}+3{\rm e}^{-K})(1+\tanh^3 K)^n+(3{\rm e}^{K}+{\rm e}^{-3K})(1-\tanh^3 K)^n
  \right].
\end{eqnarray}
This is eq. (\ref{AAresult}).


\vspace{1cm}
\hrule
\vspace{2mm}
The present version includes a few corrections in the notation
and supersedes the published version,
J. Phys. Soc. Jpn. {\bf 75} (2006) 034004.

\end{document}